\documentclass[superscriptaddress,twocolumn,showpacs,prl,floatfix]{revtex4}

\begin{document}

\title{Comment on "The Spin-Glass transition of the Three-Dimensional Heisenberg Spin Glass"}

\author{I.~A.~Campbell}
\affiliation{Laboratoire des Collo\"ides, Verres et Nanomat\'eriaux, 
Universit\'e Montpellier II, 34095 Montpellier, France}

\author{H.~Kawamura}
\affiliation{Faculty of Science, University of Osaka,
 Toyonaka 560-0043, Japan}

\begin{abstract}
Campos et al. [Phys. Rev. Lett. 97 (2006) 217204] claim that in the 3d Heisenberg Spin Glass, chiral and spin sector ordering temperatures are identical. We point out that in their analysis of their numerical data key assumptions are made which are unjustified.
\end{abstract}

\maketitle


\vskip 0.5 cm

Campos {\it et al} \cite{campos:06} present numerical spin and chiral data on the $3d$ Gaussian Heisenberg spin glass (HSG), equilibrating their systems up to the impressively large size $L=32$. Their chiral correlation length ratio $\xi(L,T)/L$ curves cross at temperatures which are only weakly $L$ dependent while the spin $\xi(L,T)/L$ curves cross at progressively lower temperatures as $L$ increases. They nevertheless make the firm claim that there is a single critical temperature for spin and chiral order in the thermodynamic limit. The key assumption on which this claim is based is that the crossing temperature shifts in the spin sector are due to strong logarithmic finite size corrections to scaling.(It is not clear why equally strong corrections should not then be expected in the chiral sector also). They state that a logarithmic or quasi-logarithmic form for the corrections is justified because at $d=3$ the HSG is at or very near to its lower critical dimension (LCD). In a number of places they propose tentative analogies with a Kosterlitz-Thouless (KT) transition. 

An essential step in the arguments given by Campos {\it et al} is the statement that finite size scaling (FSS) corrections must become logarithmic (i.e. the confluent correction exponent $\omega =0$) at an LCD. It is well known that logarithmic corrections exist generically at upper critical dimensions, but no such general rule holds for LCDs.
Campos {\it et al} refer explicitly to the case of the $2d$ Heisenberg ferromagnet. This system is indeed at its LCD with $T_c=0$, but $\omega = 2$ not $0$ with scaling corrections of the form $L^{-2}$ (multiplied by further logarithmic factors) \cite{caracciolo:95,caracciolo:98}. Thus even if the $3d$ HSG turns out to be close to its LCD, which is not proven, this would give no grounds for postulating that the leading FSS corrections are quasi-logarithmic.

We can also compare the observed properties of the HSG with those of KT systems alluded to by Campos {\it et al}, even though we are not aware of a KT ordering transition having ever been reported for any $3d$ spin system. Generically, below a KT transition the whole low temperature phase can be considered critical. Hence to within FSS corrections $\xi(L,T)/L$ and the Binder parameter $g(L,T)$ should both be independent of $L$ at all temperatures below $T_c$. In the canonical $2d XY$ ferromagnet \cite{hasenbusch:05} there is a logarithmic correction to $\xi(L,T)/L$ but only at $T_c$. The $\xi(L,T)/L$ curves never cross but merge at temperatures just below $T_c$ \cite{ballesteros:00,viet}, where the FSS correction factor becomes weak and algebraic \cite{hasenbusch:05a}. 

The behavior observed in the $3d$ HSGs is radically different. The HSG spin $\xi(L,T)/L$ curves do not merge with decreasing $T$
as in KT but intersect, with crossing points shifting to lower temperatures for large $L$ \cite{hukushima:00,campos:06}. For temperatures below the crossing points the raw spin and chiral
$\xi(L,T)/L$ curves fan out with increasing $L$ \cite{lee:03,hukushima:00,campos:06} rather than becoming $L$ independent. 

The Binder parameters $g(L,T)$ in the HSG (which were not studied by Campos {\it et al}) also show behavior entirely unlike that seen in a KT system \cite{iniguez:97,hasenbusch,viet}. In particular, the spin $g(L,T)$ curves do not show any merging behavior even at lower temperatures where, in the KT picture, the FSS correction should be small.

Thus neither the LCD argument nor the KT analogy can be considered plausible justifications for postulating a logarithmic FSS correction. Even if the logarithmic correction \cite{campos:06} is taken to be purely heuristic, by inspection its introduction does not lead to a convincing scaling for a significant range of $L$. 

On the other hand the Campos {\it et al} data without the logarithmic correction hypothesis appear to be consistent with the $T_c({\rm spin}) < T_c({\rm chiral})$ prediction of the chiral driven ordering scenario for the HSGs \cite{kawamura:92,hukushima:00,kawamura:06}, the strong spin size effects then being due to spin-chirality decoupling phenomena setting in at length scales beyond a crossover length \cite{hukushima:00,kawamura:06}. Further simulations on still larger samples could definitively resolve this question.

We would like to thank A. Pelissetto and M. Hasenbusch for their helpful advice.

\end{document}